\documentclass[runningheads]{llncs}
\usepackage{graphicx}
\usepackage{caption,subcaption}
\usepackage[utf8x]{inputenc} 
\usepackage{amsmath} 
\usepackage[noabbrev]{cleveref}
\usepackage{url}
\usepackage{mathtools}
\usepackage{amssymb}
\usepackage{placeins}
\usepackage{algorithm,algorithmic}
\DeclarePairedDelimiter{\ceil}{\lceil}{\rceil}
\DeclarePairedDelimiter{\abs}{\lvert}{\rvert}

\usepackage{crunch}
\usepackage{tikz}

\begin{document}

\mainmatter  

\title{Topological Data Analysis\\ with $\epsilon$-net Induced Lazy Witness Complex}


%
%
\author{%
Naheed Anjum Arafat \inst{1},
Debabrota Basu \inst{2},
St\'ephane Bressan \inst{1}
}%
\institute{
School of Computing, National University of Singapore, Singapore \and Department of Computer Science and Engineering, Chalmers University of Technology, G\"{o}teborg, Sweden
}
%
%

\maketitle
\begin{abstract}
Topological data analysis computes and analyses topological features of the point clouds by constructing and studying a simplicial representation of the underlying topological structure. The enthusiasm that followed the initial successes of topological data analysis was curbed by the computational cost of constructing such simplicial representations. 
The lazy witness complex is a computationally feasible approximation of the underlying topological structure of a point cloud. It is built in reference to a subset of points, called landmarks, rather than considering all the points as in the \v{C}ech and Vietoris-Rips complexes. The choice and the number of landmarks dictate the effectiveness and efficiency of the approximation.

We adopt the notion of $\epsilon$-cover to define $\epsilon$-net.
We prove that $\epsilon$-net, as a choice of landmarks, is an $\epsilon$-approximate representation of the point cloud and the induced lazy witness complex is a $3$-approximation of the induced Vietoris-Rips complex.
Furthermore, we propose three algorithms to construct $\epsilon$-net landmarks. We establish the relationship of these algorithms with the existing landmark selection algorithms. We empirically validate our theoretical claims. We empirically and comparatively evaluate the effectiveness, efficiency, and stability of the proposed algorithms on synthetic and real datasets.
\end{abstract}
\section{Introduction}
Topological data analysis computes and analyses topological features of generally high-dimensional and possibly noisy data sets. Topological data analysis is applied to eclectic domains namely shape analysis~\cite{collins2004barcode}, images~\cite{Carlsson2008,letscher2007image}, sensor network analysis~\cite{de2007coverage}, social network analysis~\cite{Carstens2013,petri2013topological,sizemore2017classification,patania2017shape}, computational neuroscience~\cite{lee2011discriminative}, and protein structure study~\cite{Xia2014PersistentFolding,kovacev2016using}.

The enthusiasm that followed the initial successes of topological data analysis was curbed by the computational challenges posed by the construction of an exact simplicial representation, the \v{C}ech complex, of the point cloud. A simplicial representation facilitates computation of basic topological objects such as simplicial complexes, filtrations, and persistent homologies. Thus, researchers devised approximations of the \v{C}ech complex as well as its best possible approximation the Vietoris-Rips complex~\cite{de2004topological,sparse_rips,gic}.
One of such computationally feasible approximate simplicial representation is the \emph{lazy witness complex}~\cite{de2004topological}. The lazy witness complex is built in reference to a subset of points, called \emph{landmarks}. The choice and the number of landmarks dictate the effectiveness and efficiency of the approximation.

We adopt the notion of $\epsilon$-cover~\cite{heinonen2012lectures} from analysis to define and present the notions of $\epsilon$-sample, $\epsilon$-sparsity, and $\epsilon$-net (\Cref{sec:net}) to capture bounds on the loss of topological features induced by the choice of landmarks. 
We prove that an $\epsilon$-net is an $\epsilon$-approximate representation of the point cloud with respect to the Hausdorff distance. We prove that the lazy witness complex induced by an $\epsilon$-net, as a choice of landmarks, is a $3$-approximation of the induced Vietoris-Rips complex.
$\epsilon$-net allows us to provide such approximation guarantees for lazy witness complex (\Cref{subsec:prop}) which was absent in the literature.
Furthermore, we propose three algorithms to construct $\epsilon$-net as landmarks for point clouds (\Cref{sec:construction}). We establish their relationship with the existing landmark selection algorithms, namely \emph{random} and \emph{maxmin}~\cite{de2004topological}. We empirically and comparatively show that the size of the \emph{$\epsilon$-net landmarks} constructed by the algorithms varies inversely with $\epsilon$ and agrees with the known bound on the size of $\epsilon$-net~\cite{krauthgamer2004navigating}. 

We empirically and comparatively validate our claim on the topological approximation quality of the lazy witness complex induced by the $\epsilon$-net landmarks (\Cref{sec:experiment}). Furthermore, we empirically and comparatively validate the effectiveness, efficiency and stability of the proposed algorithm on representative synthetic point clouds as well as a real dataset. 
Experiments confirm our claims by showing equivalent effectiveness of the algorithms constructing $\epsilon$-net landmarks with the existing maxmin algorithm. We also show the $\epsilon$-net landmarks to be more stable than those selected by the algorithms maxmin and random as $\epsilon$-net incurs narrower confidence band in the persistent landscape topological descriptor. We conclude (\Cref{sec:conc}) with the theoretical and experimental pieces of evidence that validate the $\epsilon$-nets constructed as a stable and effective way to construct landmarks and to induce lazy witness complexes.
\section{Related Works}\label{sec:related}
\textbf{Applications of TDA.} TDA is applied in different domains mostly on relatively small datasets and up to dimension $2$ due to computational intractability of the popular \v Cech and Vietoris-Rips complexes. 
\cite{collins2004barcode} computed homology classes at dimension $0$ for their proposed tangential filtration of point clouds of handwritten digits for image classification (dataset size $\sim$69–294). \cite{skraba2010persistence} used the persistence pairs at dimension $0$ for segmenting mesh on benchmark mesh segmentation datasets (size $\sim$50000). 
Researchers applying TDA to network analysis focus on characterising networks using features computed from persistence homology classes. \cite{Carstens2013} and \cite{petri2013topological} computed persistence homology at dimension 0,1 and 2 of the clique filtration to study weighted collaboration networks (size $\sim$36000) and weighted networks from different domains (size $\sim$54000) respectively. 
In biological networks, \cite{duman2018gene} clustered gene co-expression networks (size $\sim400$) based on distances between Vietoris-Rips persistence diagram computed on each network. 
In molecular biology, persistent homology reveals different conformations of proteins~\cite{Xia2014PersistentFolding,kovacev2016using} based on the strength of the bonds of the molecules. 


\textbf{Topological Approximation.} 
Computational infeasibility of the \v Cech complex and Vietoris-Rips complex motivates the development of approximate simplicial representations such as the lazy witness complexes, sparse-Rips complex~\cite{sparse_rips} and graph induced complex (GIC)~\cite{gic}.  
Sparse-Rips complex~\cite{sparse_rips} perturbs the distance metric in such a way that when the regions covered by a point can be covered by its neighbouring points, that point can be deleted without changing the topology. 
Given a graph constructed on a point cloud as input, the graph induced complex is a simplicial complex built on a subset of vertices, where the vertices of a $k$-simplex are the nearest neighbours of a clique-vertices of a $k$-clique~\cite{gic}. Due to their computational benefits, lazy witness complex and graph induced complexes have found applications in studying natural image statistics~\cite{de2004topological} and image classification~\cite{dey2017improved}.


\textbf{Applications of $\epsilon$-net.} The concept of $\epsilon$-net is a standard concept in analysis and topology~\cite{heinonen2012lectures} originating from the idea of $(\delta, \epsilon)$-limits formulated by Cauchy. 
Nets have been used in nearest-neighbor search~\cite{krauthgamer2004navigating}.~\cite{har2006fast} proposed the Net-tree data structure to represent $\epsilon$-nets at all scales of $\epsilon$. Net-tree is used to construct approximate well-separated pair decompositions~\cite{har2006fast} and approximate geometric spanners~\cite{har2006fast}. The simplicial complexes in the graph induced complex are nets. Sparse-Rips filtration constructs a net-tree on the point-cloud to decide which neighbouring points to delete. \cite{guibas2008reconstruction} used $\epsilon$-net for manifold reconstruction. 

\section{Topological Data Analysis}\label{sec:tda}
\begin{figure}[t!]
    \centering
    \includegraphics[width=0.8\textwidth]{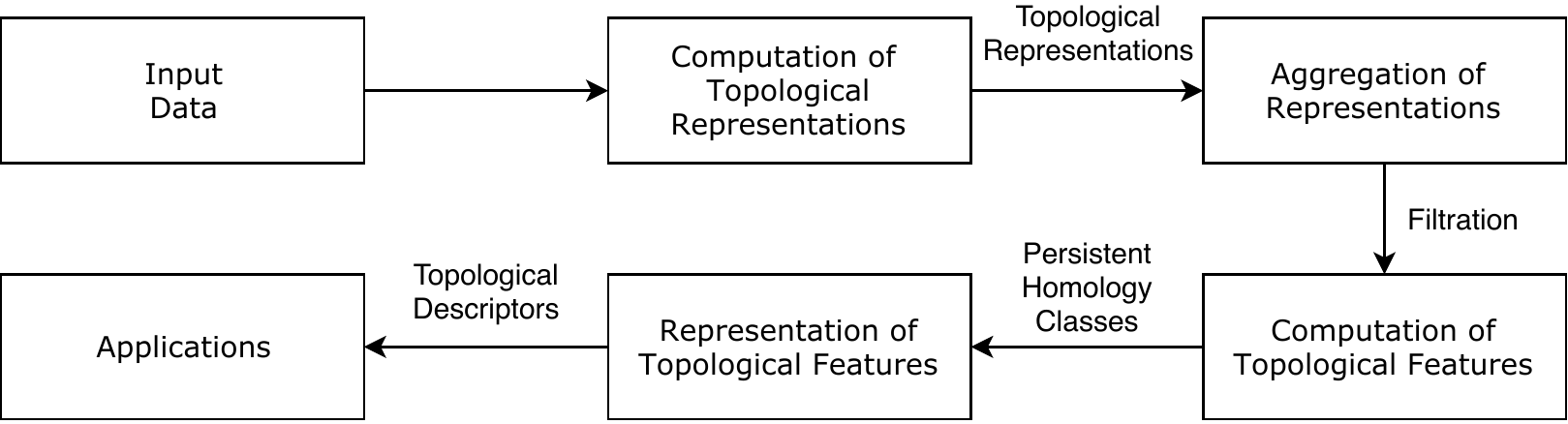}
    \caption{Components of topological data analysis.}
    \label{fig:pipeline}
\end{figure}

Topological data analysis is the study of computational models for efficient and effective computation of topological features, such as persistent homology classes, from different datasets, and representation of the topological features using different topological descriptors, such as persistence barcodes, for further analysis and application~\cite{edelsbrunner2010computational,otter2017roadmap}.
In this section, we represent the computational blocks of topological data analysis in Figure~\ref{fig:pipeline} and further describe each of the blocks.

Topological data analysis computes the topological features, such as persistent homology classes, by computing the topological objects called \emph{simplicial complex} for a given dataset. A \textbf{simplicial complex} is constructed using simplices. Formally, a \emph{$k$-simplex} is the convex-hull of $(k+1)$ data points. For instance, A 0-simplex $[v_0]$ is a single point, a 1-simplex $[v_0v_1]$ an edge, and a 2-simplex $[v_0v_1v_2]$ a filled triangle. 
A \textbf{$k$-homology class} is an equivalent class of such $k$-simplicial complexes that cannot be reduced to a lower dimensional simplicial complex~\cite{edelsbrunner2010computational}.
In order to compute the $k$-homology classes, a practitioner does not have direct access to the underlying space of the point cloud and it is combinatorially hard to compute the exact simplicial representation of \v{C}ech complex~\cite{zomorodian2010fast}.
Thus, different approximations of the exact simplicial representation are proposed: \emph{Vietoris-Rips} complex~\cite{hausmann1995vietoris} and \emph{lazy witness} complex~\cite{de2004topological}.

The \textbf{Vietoris-Rips complex} $R_\alpha(D)$, for a given dataset $D$ and real number $\alpha > 0$, is an abstract simplicial complex representation consisting of such $k$-simplices, where any two points $v_i, v_j$ in any of these $k$-simplices are at distance at most $\alpha$. Vietoris-Rips complex is the best possible ($\sqrt{2}$-)approximation of the \v{C}ech complex, computable with present computational resources, and is extensively used in topological data analysis literature~\cite{otter2017roadmap}. Thus, we use the Vietoris-Rips complex as the baseline representation in this paper.
In the worst case, the number of simplices in the Vietoris-Rips complex grows exponentially with the number of data points~\cite{otter2017roadmap,zomorodian2010fast}. 
Lazy witness complex~\cite{de2004topological} approximates the Vietoris-Rips complex by constructing the simplicial complexes over a subset of data points $L$, referred to as the landmarks. 
\begin{definition}[Lazy Witness Complex~\cite{de2004topological}]
Given a positive integer $\nu$ and real number $\alpha>0$, the \textbf{lazy witness complex} $LW_{\alpha}(D,L,\nu)$ of a dataset $D$ is a simplicial complex over a landmark set $L$ where for any two points $v_i,v_j$ of a $k$-simplex $[v_0 v_1\cdots v_k]$, there is a point $w$ whose $(d^\nu(w) + \alpha)$-neighborhood\footnote{$d^\nu(w)$ is the  distance from point $w \in D$ to its $\nu$-th nearest point in $L$.} contains $v_i, v_j$.
\end{definition}
In the worst case, the size of the lazy witness complexes grows exponentially with the number of landmarks. Less number of landmarks facilitates computational acceleration while produces a bad approximation of Vietoris-Rips with loss of topological features. Thus, the trade-off between the approximation of topological features and available computational resources dictates the choice of landmarks. 

As the value of filtration parameter $\alpha$ increases, new simplices arrive and the topological features, i.e. the homology classes, start to appear. Some of the homology classes merge with the existing classes in a subsequent simplicial complex, and some of them persist indefinitely~\cite{edelsbrunner2010computational}.
In order to capture the evolution of topological structure with scale, topological data analysis techniques construct a sequence of simplicial complex representations, called a \emph{filtration}~\cite{edelsbrunner2010computational}, for an increasing sequence of $\alpha$'s. 
In a given filtration, the persistence interval of a homology class is denoted by $[\alpha_{b},\alpha_{d})$, where $\alpha_{b}$ and $\alpha_{d}$ are the filtration values of its appearance and merging respectively. The persistence interval of an indefinitely persisting homology class is denoted as $[\alpha_{b},\infty)$.

Topological descriptors, such as persistence diagram~\cite{edelsbrunner2010computational} and persistence landscapes~\cite{bubenik2015statistical} represent persistence intervals as points and functions respectively in order to draw qualitative and quantitative inference about the topological features. Distance measures between persistent diagrams such as the bottleneck and Wasserstein distances  ~\cite{edelsbrunner2010computational} are often used to draw quantitative inference. The bottleneck distance between two diagrams is the smallest distance $d$ for which there is a perfect matching between the points of the two diagrams such that any pair of matched points are at distance at most $d$~\cite{edelsbrunner2010computational}. The Wasserstein distance between two diagrams is the cost of the optimal matching between points of the two diagrams~\cite{edelsbrunner2010computational}.
\section{$\epsilon$-net}
\label{sec:net}
As we discussed in \Cref{sec:tda}, topological data analysis of a dataset begins with the computation of simplicial complex representations. 
Though Vietoris-Rips is the best possible approximation of the \v Cech complex, it incurs an exponential computational cost with respect to the size of the point cloud. Thus, lazy witness complex is often used as a practical solution for scalable topological data analysis.
Computation of lazy witness complex is dependent on the selection of landmarks.
Selection of landmarks dictates the trade-off between \textit{effectiveness}, i.e. the quality of approximation of topological features, and \textit{efficiency}, i.e. the computational cost of computing the lazy witness complex. 

$\epsilon$-cover is a construction used in topology to compute the inherent properties of a given space~\cite{heinonen2012lectures}.
In this paper, we import the concept of $\epsilon$-cover to define $\epsilon$-net of a point cloud. We use the $\epsilon$-net of the point cloud as the landmarks for constructing lazy witness complex. We show that $\epsilon$-net, as a choice of landmarks, has guarantees such as being an $\epsilon$-approximate representation of the point cloud, its induced lazy witness complex being a $3$-approximation of its induced Vietoris-Rips complex, and also bounding the number of landmarks for a given $\epsilon$. These guarantees are absent for the other existing landmark selection algorithms (Section~\ref{sec:construction}) such as random and maxmin algorithms.

\subsection{$\epsilon$-net of a Point Cloud}
In this section, we define the $\epsilon$-net of a point cloud in reference to the $\epsilon$-cover used in topology. 
$\epsilon$-cover is a set of subsets of a point cloud in an Euclidean space such that these subsets together cover the point cloud, but none of the subsets has a diameter more than $\epsilon$.
\begin{definition}[$\epsilon$-cover~\cite{heinonen2012lectures}]
An $\epsilon$-cover of a point cloud $P$ is the set of $P_i$'s such that $P_i \subseteq P$, $P = \cup_{i} P_i$, and diameter\footnote{The diameter $diam(P_i)$ of a set $P_i \subseteq P$ is defined as the largest distance $d(x,y)$ between any two points in $x, y \in P_i$.} of $P_i$ is at most $\epsilon \geq 0$ for all $i$.
\end{definition}
When the sets in the $2\epsilon$-cover of $P$ are Euclidean balls of radius $\epsilon$, the set of centres of the balls is termed as an $\epsilon$-sample of set $P$.\vspace*{-.5em}
\begin{definition}[$\epsilon$-sample~\cite{guibas2008reconstruction}]
A set $L \subseteq P$ is an $\epsilon$-sample of $P$ if the collection $\{B_{\epsilon}(x):x\in L\}$ of $\epsilon$-balls of radius $\epsilon$ covers $P$ , i.e. $P = \cup_{x \in L} B_\epsilon(x)$.
\end{definition}
According to the definition of $\epsilon$-sample, $P$ is an $\epsilon$-sample of itself for $\epsilon = 0$. For decreasing further computational expense, it is desirable to have an $\epsilon$-sample is sparse that means it contains as less number of points as possible. An $\epsilon$-sparse subset of $P$ is a subset where any two points are at least $\epsilon$ apart from each other.
\begin{definition}[$\epsilon$-sparse]
A set $L \subset P$ is $\epsilon$-sparse if for all $x,y \in L$, $d(x,y) > \epsilon$.
\end{definition}
An $\epsilon$-net of set $P$ is an $\epsilon$-sparse subset of $P$ which is also an $\epsilon$-sample of $P$.\vspace*{-.5em}
\begin{definition}[$\epsilon$-net~\cite{heinonen2012lectures}] 
Let $(P,d)$ be a metric space and $\epsilon \geq 0$. A subset $L \subset P$ is called an $\epsilon$-net of $P$ if $L$ is  $\epsilon$-sparse and an $\epsilon$-sample of $P$.
\end{definition}

\subsection{Properties of $\epsilon$-nets}
\label{subsec:prop}
$\epsilon$-net of a point cloud comes with approximation guarantees irrespective of its algorithmic construction. An $\epsilon$-net of a point cloud of diameter $\Delta$ in Euclidean space $\mathbb{R}^D$ is an $\epsilon$-approximation of the point cloud in Hausdorff distance. The lazy witness complex induced by an $\epsilon$-net is a $3$-approximation of the Vietoris-Rips complex on that $\epsilon$-net. Furthermore, the size of an $\epsilon$-net is at most $(\frac{\Delta}{\epsilon})^{\theta(D)}$~\cite{krauthgamer2004navigating}. Here, we establish the first two approximation guarantees of $\epsilon$-net theoretically. In section ~\ref{sec:experiment} we validate the last two properties empirically for the proposed  algorithms for constructing $\epsilon$-nets.

\textbf{Point-cloud Approximation Guarantee of an $\epsilon$-net.}\label{subsec:hausdorff}
We use \Cref{lemma:esample} to prove that the $\epsilon$-net of a point cloud $P$ is an $\epsilon$-approximate representation of that point cloud in Hausdorff distance. 
\begin{lemma}\label{lemma:esample}
Let $L$ be an $\epsilon$-net of point cloud $P$. For any point $p \in P$, there exists a point $q \in L$ such that the distance $d(p,q) \leq \epsilon$
\end{lemma}
\begin{theorem}
\label{thm:hausdorff}
The Hausdorff distance between the point cloud $P$ and its  $\epsilon$-net $L \subseteq P$ is at most $\epsilon$.
\end{theorem}
\begin{proof}
For any $l \in L$, there exists a point $p \in P$ such that $d(l,p) \leq \epsilon$, by definition of $B_{\epsilon}(l)$. Hence, $\min_{l \in L}{d(l,p)} \leq \epsilon$, and thus, $\max_{p \in P}{\min_{l \in L}{d(l,v)}} \leq \epsilon$. For any $p \in P$, there exists a landmark $l \in L$ such that $d(l,p) \leq \epsilon$, by Lemma~\ref{lemma:esample}. Thus, $\max_{l \in L}{\min_{p \in P}{d(l,p)}} \leq \epsilon$. Hence the Hausdorff distance $d_H(P,L)$ between $P$ and $L$, defined as the maximum of $\max_{l \in L} \min_{p \in P}{d(l,p)}$ and $\max_{p \in P}{\min_{l \in L}{d(l,p)}}$   is bounded by $\epsilon$.
\end{proof}
\textbf{Topological Approximation Guarantee of an $\epsilon$-net Induced Lazy Witness Complex.} 
\label{subsec:theory}
In addition to an $\epsilon$-net being an $\epsilon$-approximation of the point-cloud, we prove that the lazy witness complex induced by the $\epsilon$-net landmarks is a good approximation (\Cref{thm:approx}) to the Vietoris-Rips complex on the landmarks. This approximation ratio is independent of the algorithm constructing the $\epsilon$-net. As a step towards \Cref{thm:approx}, we state \Cref{lemma:nn1} that follows from the definition of the lazy witness complex and $\epsilon$-sample. \Cref{lemma:nn1} establishes the relation between 1-nearest neighbour of points in an $\epsilon$-net.
\begin{lemma}\label{lemma:nn1}
If $L$ is an $\epsilon$-net landmark of point cloud $P$, then the distance $d(p,p')$ from any point $p \in P$ to its 1-nearest neighbour $p' \in P$ is at most $\epsilon$.
\end{lemma}
\Cref{thm:approx} shows that the lazy witness complex induced by the landmarks in an $\epsilon$-net is a $3$-approximation of the Vietoris-Rips complex on those landmarks above the value $2\epsilon$ of filtration parameter. 
\begin{theorem}
\label{thm:approx}
	If $L$ is an $\epsilon$-net of the point cloud $P$ for $\epsilon \in \mathbb{R}^+$, $LW_{\alpha}(P,L,\nu = 1)$ is the lazy witness complex of $L$ at filtration value $\alpha$, and $R_\alpha(L)$ is the Vietoris-Rips complex of $L$ at filtration $\alpha$, then
	$R_{\alpha/3}(L) \subseteq LW_\alpha(P,L,1) \subseteq R_{3\alpha}(L)$ for $\alpha \geq 2\epsilon$.
\end{theorem}
\begin{proof}	
In order to prove the first inclusion, consider a $k$-simplex $\sigma_k = [x_0x_1 \cdots x_k] \in R_{\alpha/3}(L)$. For any edge $[x_ix_j] \in \sigma_k$, let $w_t$ be the point in $P$ that is nearest to the vertices of $[x_ix_j]$ and wlog, let the point corresponding to that vertex be $x_j$. 
	Since $w_t$ is the nearest neighbour of $x_j$, by Lemma~\ref{lemma:nn1}, $d(w_t,x_j) \leq \epsilon \leq \frac{\alpha}{2}$. 
	Since $[x_ix_j] \in  R_{\alpha/3}$, $d(x_i,x_j) \leq \frac{\alpha}{3} < \frac{\alpha}{2}$. By triangle inequality, $d(w_t,x_i) \leq \frac{\alpha}{2} +  \frac{\alpha}{2} \leq \alpha$. Hence, $x_i$ is within distance $\alpha$ from $w_t$. The $\alpha$-neighbourhood of point $w_t$ contains both $x_i$ and $x_j$. 
	Since $d^1(w_t) > 0$, the $(d^1(w_t) + \alpha)$-neighbourhood of $w_t$ also contains $x_i,x_j$. Therefore, $[x_ix_j]$ is an edge in $LW_\alpha(P,L,1)$. Since the argument is true for any $x_i,x_j \in \sigma_k$, the $k$-simplex $\sigma_k \in LW_\alpha(P,L,1)$.
	
	In order to prove the second inclusion, consider a $k$-simplex $\sigma_k = [x_0x_1 \cdots x_k] \in LW_\alpha(P,L,1)$. Therefore, by definition of lazy witness complex, for any edge $[x_ix_j]$ of $\sigma_k$ there is a witness $w \in P$ such that, the $(d^1(w) + \alpha)$-neighbourhood of $w$ contains both $x_i$ and $x_j$. Hence, $d(w,x_i) \leq d^1(w) + \alpha \leq \epsilon + \alpha$  (by Lemma ~\ref{lemma:nn1})$\leq 3\alpha/2$. Similarly, $d(w,x_j) \leq 3\alpha/2$. By triangle inequality, $d(x_i,x_j) \leq 3\alpha$. Therefore, $[x_ix_j]$ is an edge in $R_{3\alpha}(L)$. Since the argument is true for any  $x_i,x_j \in \sigma_k$, the k-simplex $\sigma_k \in R_{3\alpha}(L)$.  
\end{proof}

\emph{Discussion.} Theorem~\ref{thm:approx} implies that the interleaving of lazy witness filtration $LW = {LW_\alpha(L)}$ and the Vietoris-Rips filtration $R = R_\alpha(L)$ occurs when $\alpha > 2\epsilon$. As a consequence, the weak-stability theorem~\cite{chazal2009proximity} implies that the bottleneck distance between the partial persistence diagrams $Dgm_{>2\epsilon}(LW)$ and $Dgm_{>2\epsilon}(R)$ is upper bounded by $3\log{3}$. In \Cref{sec:experiment}, we empirically validate this bound.

\textbf{Size of an $\epsilon$-net.}
The size of an $\epsilon$-net depends on $\epsilon$, the diameter of the point-cloud and the dimension of the underlying space~\cite{krauthgamer2004navigating,har2006fast}. If a point cloud $P \subset \mathbb{R}^D$ has diameter $\Delta$, the size of an $\epsilon$-net of $P$ is $(\frac{\Delta}{\epsilon})^{\theta(D)}$~\cite{krauthgamer2004navigating}. The size of an $\epsilon$-net does not depend on the size of the point cloud. In Section~\ref{sec:experiment}, we empirically validate this bound for the $\epsilon$-net landmarks generated by the proposed algorithms. The framework of $\epsilon$-net along with its approximation guarantees lead to the question of its algorithmic construction as landmarks.

\section{Construction of an $\epsilon$-net}
\label{sec:construction}
The na\"ive algorithm~\cite{har2015net} to construct an $\epsilon$-net selects the first point $l_1$ uniformly at random. In $i$-th iteration, it marks the points at a distance less than $\epsilon$ from the previously selected landmark $l_{i-1}$ as covered, and selects the new point $l_{i}$ from the unmarked points arbitrarily until all points are marked~\cite{har2006fast}. The fundamental principle is to choose, at each iteration, a new landmark from the set of yet-to-cover points such that it retains the $\epsilon$-net property. We propose three algorithms where this choice determines the algorithm. 
\subsection{Three Algorithms: $\epsilon$-net-rand, $\epsilon$-net-maxmin, and $(\epsilon,2\epsilon)$-net}
The algorithm \textbf{$\epsilon$-net-rand}, at each iteration, marks the points at a distance less than $\epsilon$ from the previously chosen landmark as covered
and chooses a new landmark uniformly at random from the unmarked points. 
The algorithm \textbf{$\epsilon$-net-maxmin}, at each iteration, marks the points at a distance less than $\epsilon$ from the previously chosen landmark as covered
and chooses the farthest unmarked point from the already chosen landmarks. It terminates when the distance to the farthest unmarked point is no more than $\epsilon$. 
The algorithm \textbf{$(\epsilon,2\epsilon)$-net}, at each iteration, marks the points at a distance less than $\epsilon$ from the previously chosen landmark as covered, and chooses a landmark uniformly at random from those unmarked points whose distance to the previously chosen landmark is at most $2\epsilon$. If there are no unmarked points at a distance in-between $\epsilon$ and $2\epsilon$ from the previous landmark, it searches for unmarked points at a distance between  $2\epsilon$ and $4\epsilon$, $4\epsilon$ and $8\epsilon$, and so on, until it either finds one to continue as before or all points are marked.
The pseudo-code for $(\epsilon,2\epsilon)$-net is in Algorithm~\ref{alg:1}. 

$(\epsilon,2\epsilon)$-net attempts to cover the point-cloud with intersecting balls of radius $\epsilon$, whereas $\epsilon$-net-maxmin attempts to cover the point-cloud with non-intersecting balls of radius $\epsilon$. $\epsilon$-net-rand does not maintain any invariant. 

 $\epsilon$-net-rand and $(\epsilon,2\epsilon)$-net have the time-complexity of $O(\frac{1}{\epsilon^D})$ and $O(\frac{1}{\epsilon^D}\log(\frac{1}{\epsilon}))$ respectively. Their run-time does not depend on the size of the input point cloud. On the other hand, the run-time of $\epsilon$-net-maxmin depends on the size of the point-cloud as it has to search for the farthest point from the landmarks at each iteration. On a point cloud of sinze $n$, $\epsilon$-net-maxmin has $O(\frac{n}{\epsilon^D})$ time-complexity.
\subsection{Connecting $\epsilon$-net to Random and Maxmin algorithms}
De Silva et al.\cite{de2004topological} proposed random and maxmin algorithms for point clouds.
\setlength{\textfloatsep}{2pt}
\begin{algorithm}[t!]
\caption{Algorithm $(\epsilon,2\epsilon)$-net}
\begin{algorithmic}[1]
\label{alg:1}
\REQUIRE Point cloud $P = \{p_1,p_2,\cdots,p_n\}$, $n \times n$ Distance matrix $D$, parameter $\epsilon$.
\ENSURE Set of Landmarks $L$.

\STATE Select the initial landmark $l_1$ uniformly at random from $P$.
\STATE Initialize $L = \{l_1\}$.
\STATE Let $N^1_{(\epsilon,2\epsilon)}$ be the set of points at a distance between $\epsilon$ and $2\epsilon$  from $l_1$.
\STATE Initialize candidate landmarks $C_1 = N^1_{(\epsilon,2\epsilon)}$.
\STATE $i = 1$.
 \REPEAT 
    \STATE Let $N^i_{\leq\epsilon}$ be the set of points at a distance less than $\epsilon$ from $l_i$.
    \STATE Mark all the points in $N^i_{\leq\epsilon}$ as covered.
    \STATE Let $C_i^{u}$  be the set of unmarked points in $C_i$.
    \IF{$C_i^u$ is empty}
    \STATE Find the first $\delta =   [1,2,\cdots,\log(\ceil{\Delta/2\epsilon})]$ for which $N^i_{(2^\delta\epsilon,2^{\delta+1}\epsilon)}$ contains any unmarked point.
    \STATE Set $C_i = C_i \cup  N^i_{(2^\delta\epsilon,2^{\delta+1}\epsilon)}$.
    \ENDIF
    \STATE Select $l_{i+1}$ uniformly at random from $C_i^{u}$.
    \STATE Insert $l_{i+1}$ to L.
     \STATE $C_{i+1} = C_i \cup N^{i+1}_{(\epsilon,2\epsilon)}$.
    \STATE $i = i + 1$.
 \UNTIL{all points are marked}
\end{algorithmic}
\end{algorithm}

\paragraph{Random.} Given a point cloud $P$, the algorithm \emph{random} selects $|L|$ points uniformly at random from the set of points $P$.
This algorithm is closely related to $\epsilon$-nets. Given the number of landmarks $K>1$, the set of landmarks selected by random is $\delta$-sparse where $\delta$ is the minimum of the pairwise distances among the landmarks. However, the same choice of $K$ may not necessarily make the landmarks a $\delta$-sample of the point cloud. 

The $\epsilon$-net-rand algorithm is a modification of random that takes $\epsilon$ as a parameter instead of $K$ and use $\epsilon$ to put a constraint on the domain of random choices. It continues to select landmarks until all points are marked to ensure the $\epsilon$-sample property. The proof sketch of the fact that the constructed landmarks are $\epsilon$-sparse and $\epsilon$-sample is as follows:

\emph{Proof of Correctness.} The $\epsilon$-net-rand algorithm does not terminate until all points are marked as covered. Hence the set of landmarks selected by $\epsilon$-net-rand is an $\epsilon$-sample, since otherwise, there would have been unmarked points. The pairwise distance between any two landmarks cannot be less than $\epsilon$; otherwise, one of them would have been marked by the other and the marked point would not be a landmark. Hence the set landmarks selected by $\epsilon$-net-rand is $\epsilon$-sparse.

\paragraph{Maxmin.} The \emph{maxmin} algorithm selects the first landmark $l_1$ uniformly at random from the set of points, $P$. Following that; it selects the point which is furthest to the present set of landmarks at each step till a given number of landmarks, say $\abs{L}$, are chosen. If $L_{i-1}=\{l_1,l_2,\dots,l_{i-1}\}$ is the set of already chosen landmarks, it selects such a point $u \in P \setminus L_{i-1}$ as the $i^{th}$ landmark that maximises the minimum distance from the present set of landmarks $L_{i-1}$. Mathematically, $l_i \triangleq {\arg\max}_{u \in P\setminus L_{i-1}}  \min_{v\in L_{i-1}} d(u,v).$
The maxmin algorithm selects landmarks such that the point cloud is covered as vastly as possible. 

The maxmin algorithm is closely related to $\epsilon$-net. Given the number of landmarks $K>1$, the set of landmarks selected by maxmin is $\delta$-sparse where $\delta$ is the minimum of the pairwise distances among the landmarks chosen. However, that choice of $K$ may not necessarily make the landmarks a $\delta$-sample of the point cloud.
The $\epsilon$-net-maxmin algorithm is a modification of maxmin that takes $\epsilon$ as a parameter instead of $K$ and uses $\epsilon$ to control sparsity among the landmarks. It terminates when the minimum of the pairwise distances among the landmarks drops below $\epsilon$ to ensure the $\epsilon$-sample property of the landmarks chosen. The proof sketch of the resulting landmarks being $\epsilon$-sparse and $\epsilon$-sample is as follows:

\emph{Proof of Correctness.} The $\epsilon$-net-maxmin algorithm, at each iteration, selects only that point as a landmark whose minimum distance to the other landmark points is the largest among all unmarked points. If such a point's minimum distance to the other landmark points is no more than $\epsilon$, the algorithm terminates. Hence the set of landmarks selected by $\epsilon$-net-maxmin must be $\epsilon$-sparse. A point that is not a landmark must be covered by some landmark point already. Otherwise, its minimum distance to the landmark set would have been at least $\epsilon$, and hence would have been the only unmarked point available to be selected as a new landmark by $\epsilon$-net-maxmin. Therefore the set of landmarks selected by $\epsilon$-net-maxmin is also $\epsilon$-sample of the point cloud.
\section{Empirical Performance Evaluation}\label{sec:experiment}
We implement the pipeline illustrated in~\Cref{fig:pipeline} to empirically validate our theoretical claims and also the effectiveness, efficiency, and stability of the algorithms that construct $\epsilon$-net landmarks compared to that of the random and maxmin algorithms. We test and evaluate these algorithms on two synthetic point cloud datasets, namely Torus and Tangled-torus, and a real-world point cloud dataset, namely 1grm. On each input point cloud, we compute the lazy witness filtration and Vietoris-Rips filtration induced by the landmarks, as well as the Vietoris-Rips filtration induced by the point cloud. 

On each dataset, as we vary parameter $\epsilon$ of the algorithms constructing $\epsilon$-nets, we study the relationship between $\epsilon$ to the number of landmarks, the quality of the topological features approximated by the lazy witness filtration induced by those landmarks, as well as the stability of those approximated features. As the algorithms maxmin and random require the number of landmarks a priori, for the sake of comparison, we use the same number of landmarks as that of the corresponding $\epsilon$-net algorithm for a given $\epsilon$.

We compute the quality of the features approximated by an algorithm in terms of the 1-Wasserstein distance between the lazy witness filtration induced by the landmarks selected by that algorithm to those of a Vietoris-Rips filtration on the same dataset. As there are elements of randomness in the algorithms, we run each experiment 10 times and compute distances averaged over the runs. 

We compute the stability of the features approximated by the algorithms in terms of the $95\%$ confidence band corresponding the rank $1$ persistence landscape using bootstrap~\cite{chazal2015subsampling}. We use persistence landscape to validate the stability of the filtrations because unlike persistence diagrams and barcodes, two sets of persistence landscapes always have unique mean and by strong law of large numbers the empirical mean landscapes of sufficiently large collection converge to its expected landscapes~\cite{bubenik2015statistical}. 

\subsection{Datasets and Experimental Setup}
\begin{figure}[t!]
\includegraphics[width=0.33\textwidth,trim=2cm 6cm 2cm 7cm]{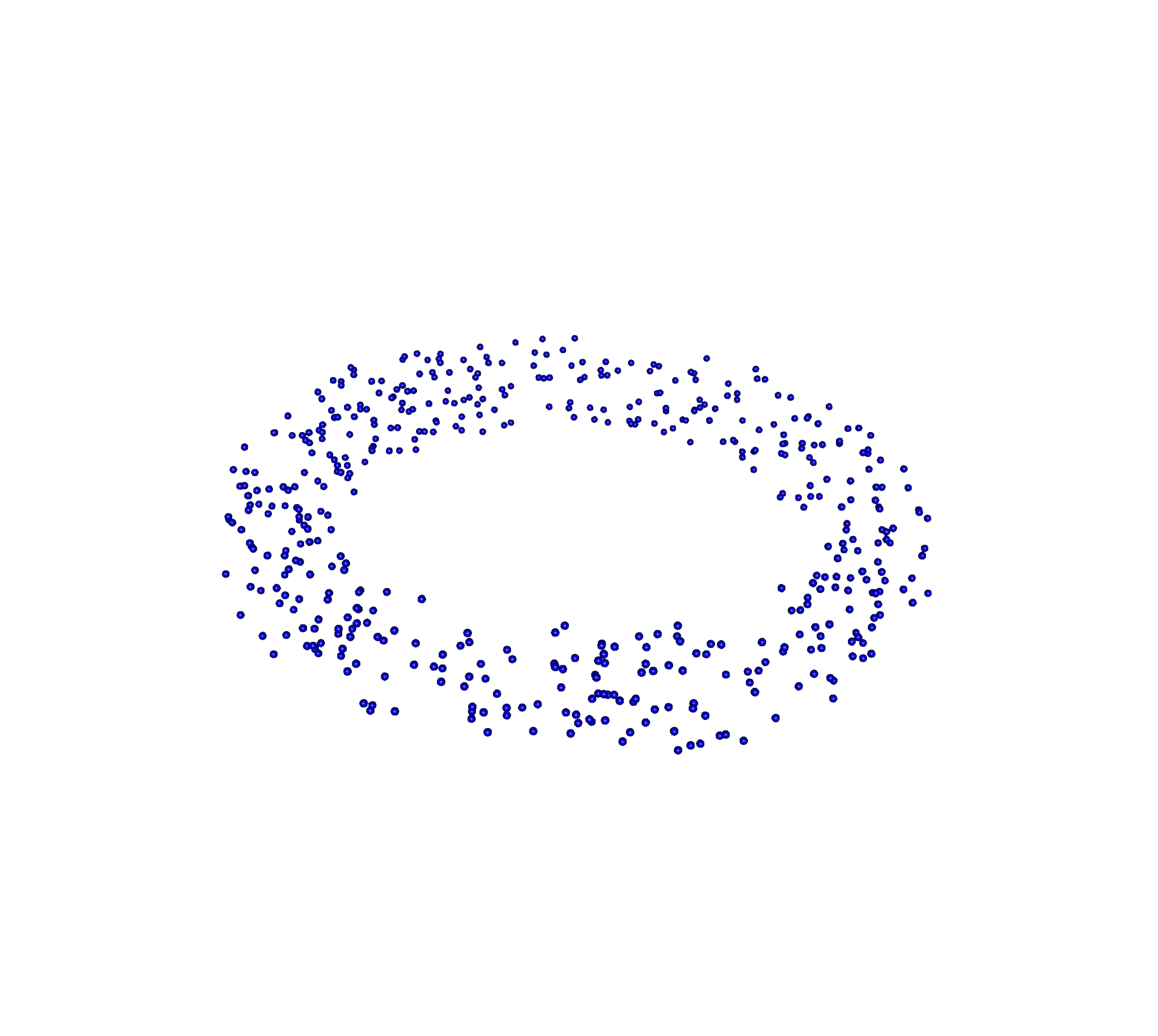}
\includegraphics[width=0.33\textwidth,trim=2cm 4cm 2cm 4cm]{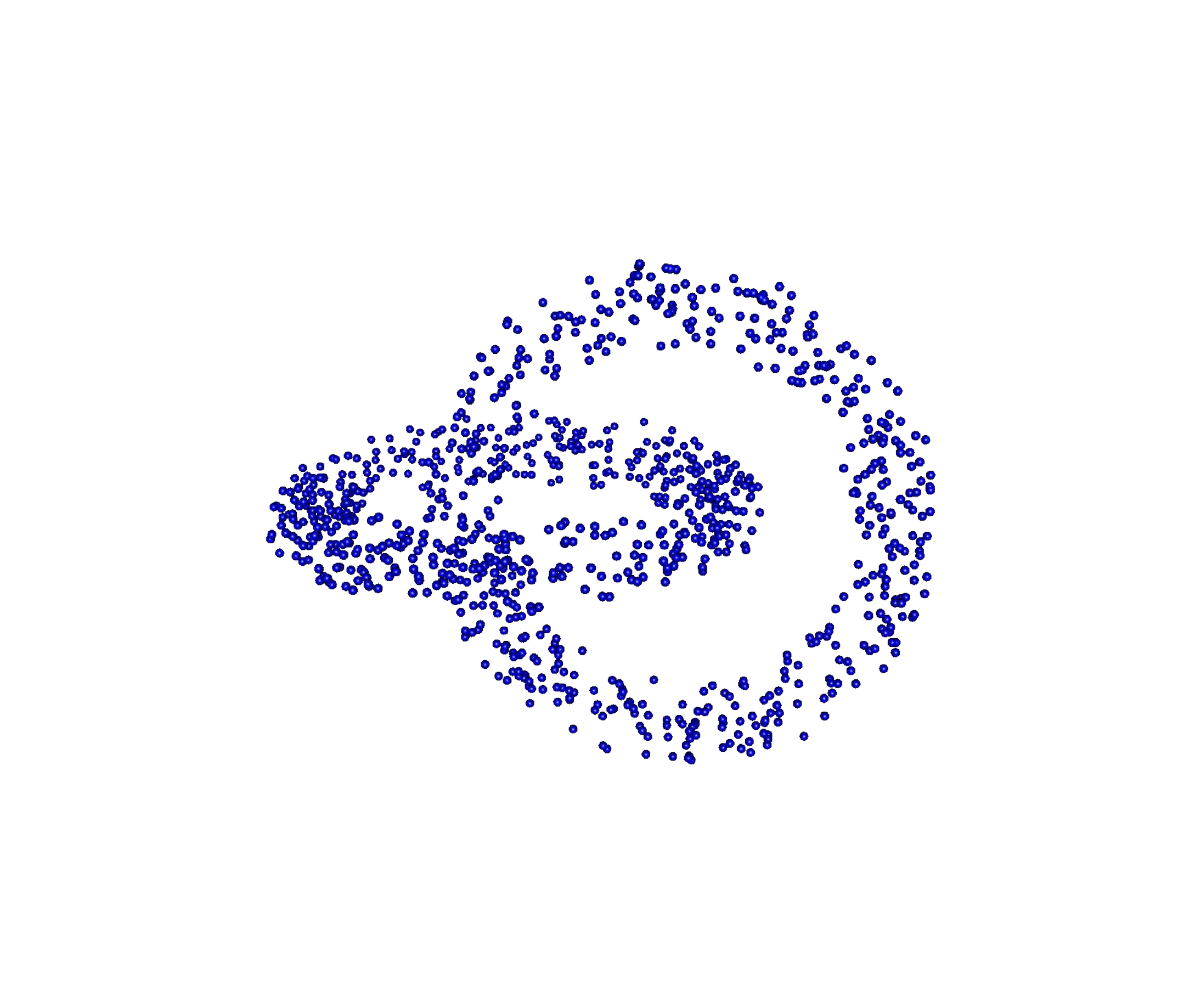}\hfill
\includegraphics[scale=0.2,trim=7cm 7cm 0cm 15cm]{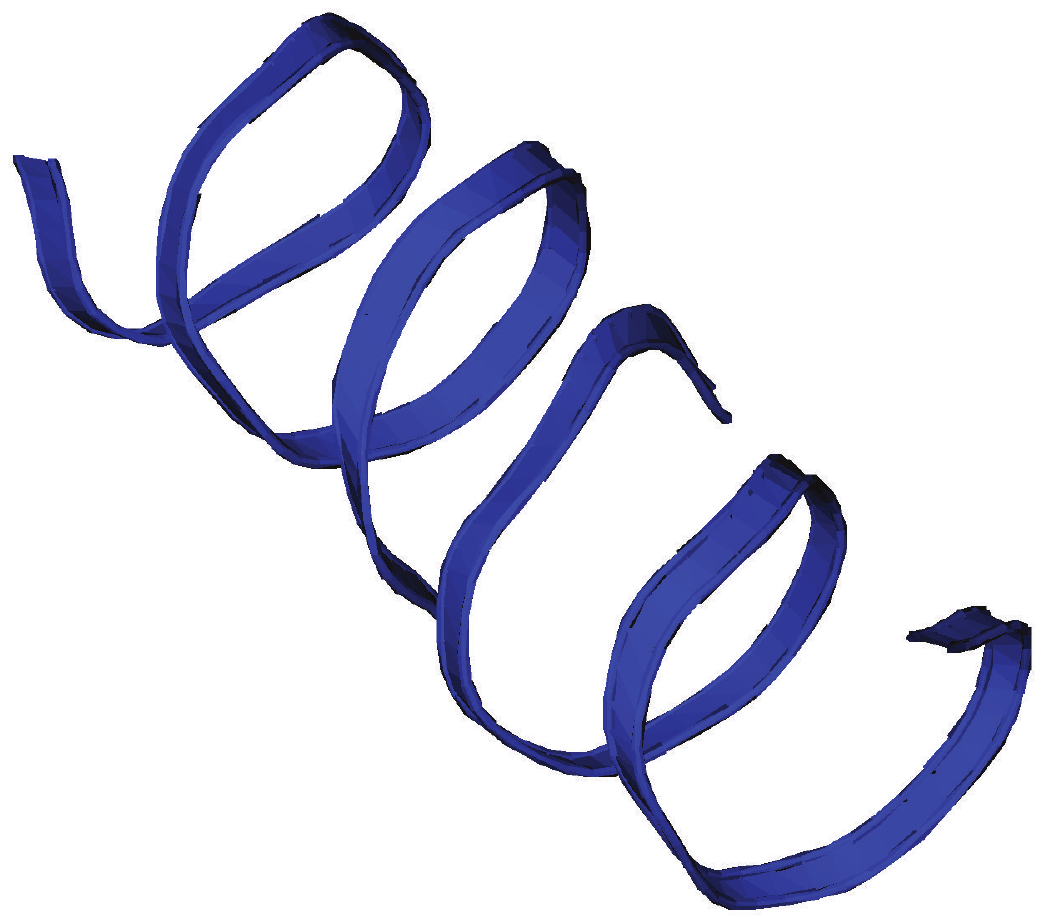}
\caption{(left) \label{fig:torus_img} Torus, (middle) \label{fig:ttorus_img} Tangled-torus, and (right) \label{fig:grm_img} 1grm Dataset}
\end{figure}

\paragraph{Datasets.}We use the datasets illustrated in Figure~\ref{fig:grm_img} for experimentation. The dataset \textbf{Torus} is a point cloud of size $500$ sampled uniformly at random from the surface of a torus in $\mathbb{R}^3$. The torus has a major radius of $2.5$ and minor radius of $0.5$. The dataset \textbf{Tangled-torus} is a point cloud of size $1000$ sampled uniformly at random from two tori tangled with each other in $\mathbf{R}^3$. Both tori has a major radius of $2.5$ and minor radius of $0.5$. The dataset \textbf{1grm} is the conformation of the gramicidin-A protein. It has a helical shape. Gramicidin-A has two disconnected chains of monomers consisting of $272$ atoms. 

\paragraph{Experimental Setup.}We implement the experimental workflow in Matlab 2018a (with 80GB memory limit).
All experiments are run on a machine with an Intel(R) Xeon(R)@2.20GHz CPU and 196 GB memory.
We use the Javaplex library~\cite{Javaplex} to construct lazy witness filtrations and to compute their persistence intervals. 
We use the Ripser library to construct the Vietoris-Rips filtrations and to compute their persistence intervals. We use R-TDA package~\cite{rTDA} to compute bottleneck and Wasserstein distances, and $95\%$ confidence band for the landscapes. We set the lazy witness parameter $\nu = 1$ in all computations.
\subsection{Validation of Theoretical Claims} 
\begin{figure}[t!]
\centering
\begin{minipage}{.45\textwidth}
  \centering
  \includegraphics[height=0.2\textheight, trim=6cm 8cm 6cm 7.5cm]{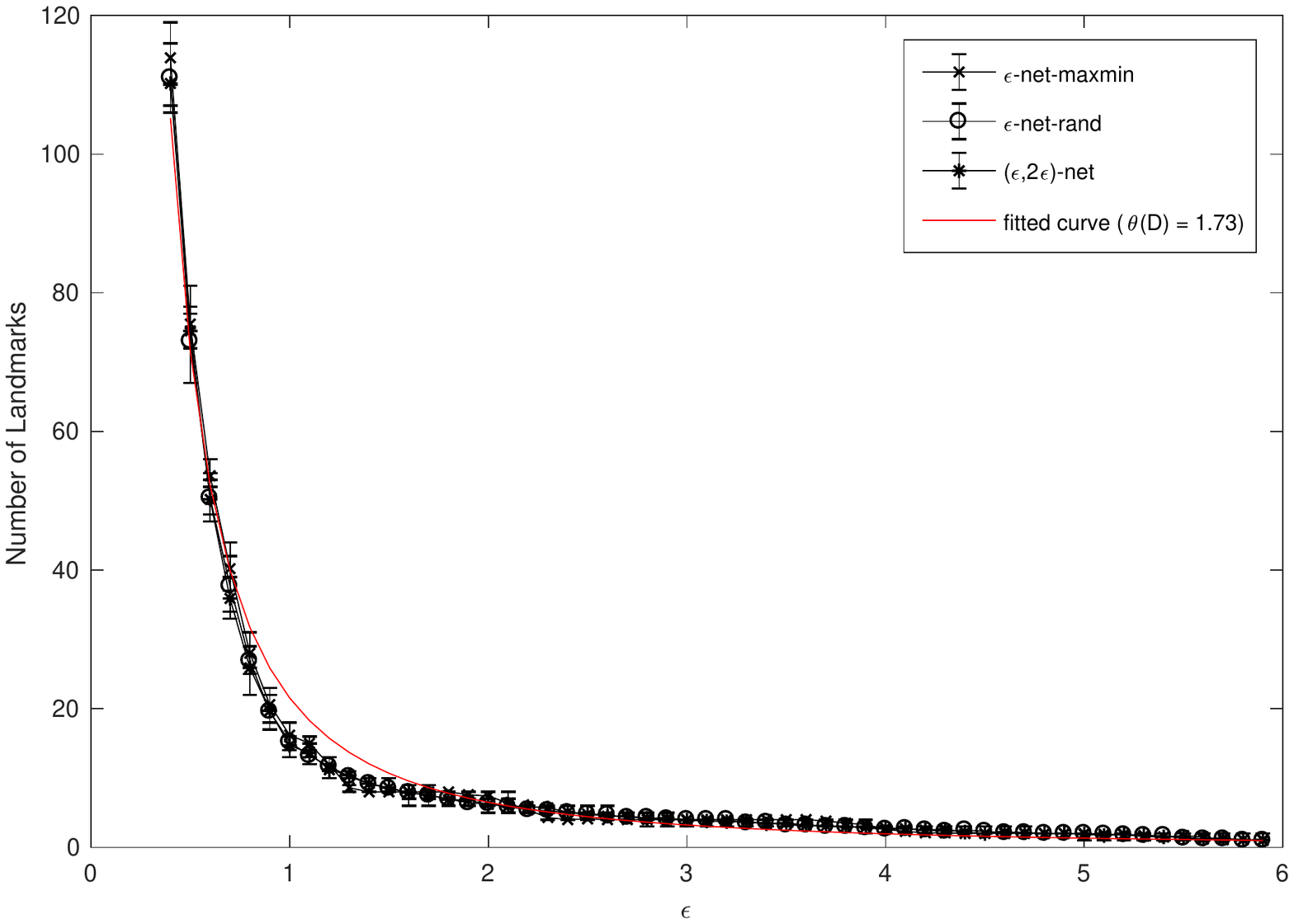}
  \captionof{figure}{\label{fig:torus_rel} Number of landmarks generated by the $\epsilon$-net algorithms vs. $\epsilon$ on Torus dataset.}
\end{minipage}\hfill
\begin{minipage}{.5\textwidth}
  \centering
\includegraphics[height=0.2\textheight, trim=3.5cm 8cm 3cm 8cm]{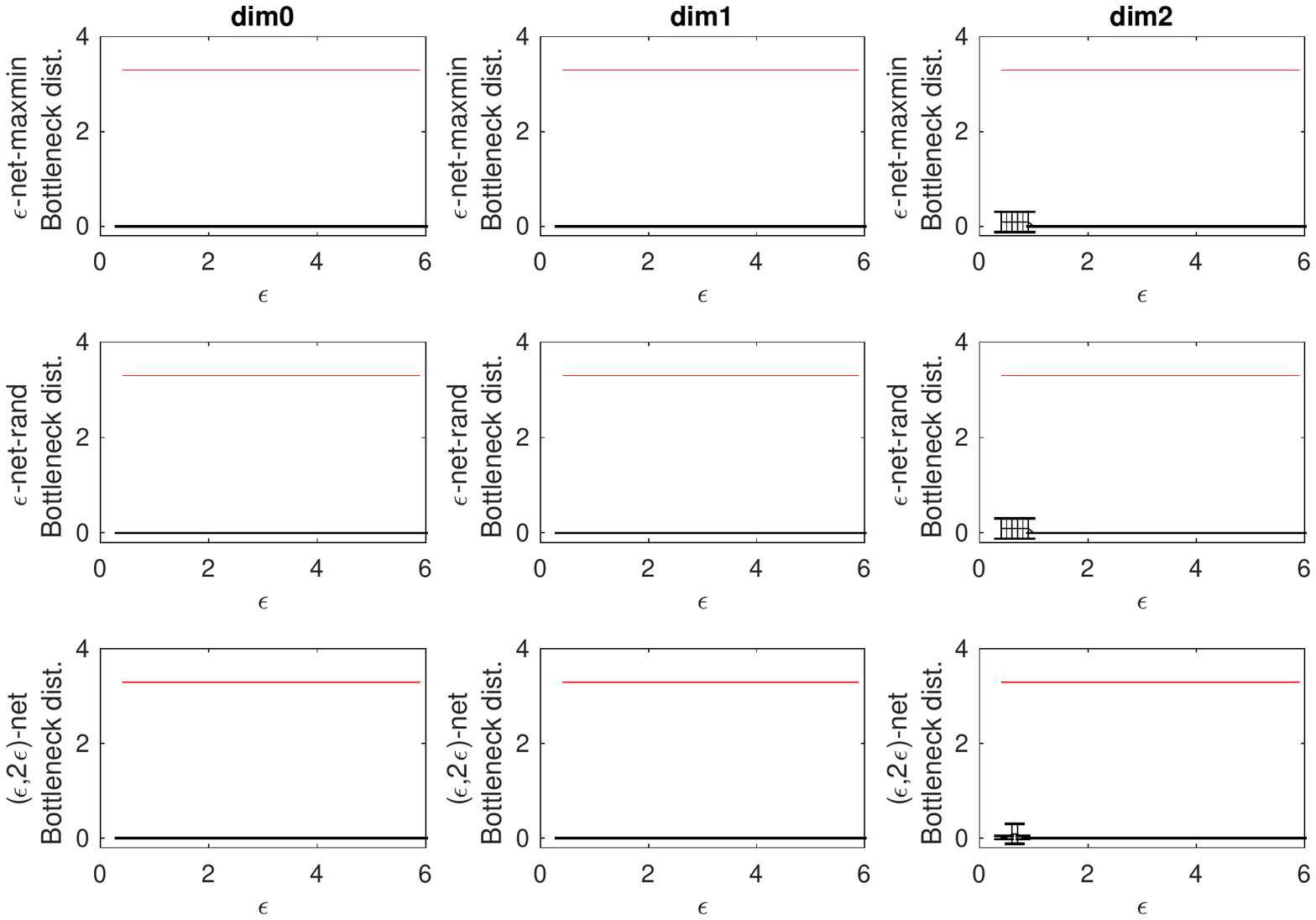}
\captionof{figure}{\label{fig:torus_vali} Topological approximation guarantee of $\epsilon$-net constructed by the algorithms on Torus dataset.}
\end{minipage}
\end{figure}

\textbf{Number of Landmarks Generated by the $\epsilon$-net Algorithms.} In Figure~\ref{fig:torus_rel}, we illustrate the relation between the number of landmarks generated by the $\epsilon$-net algorithms and $\epsilon$ on Torus dataset. Each algorithm is run $10$ times for each $\epsilon$, and the mean and standard deviation are plotted. We observe that the number of landmarks decreases as $\epsilon$ increases. We also observe that, for a fixed $\epsilon$, the average number of landmarks selected by the $\epsilon$-net algorithms is more or less stable across different algorithms. We use the number of landmarks of an $\epsilon$-net-maxmin to fit a curve with values $\Delta = 5.9$ (the diameter of Torus) and coefficient $\theta(D) = 1.73$ (found from fitting with $95\%$ confidence). This observation supports the theoretical bound of $(\frac{\Delta}{\epsilon})^{\theta(D)}$. 

\textbf{Topological Approximation Guarantee.} In order to validate Theorem~\ref{thm:approx} on dataset Torus, we compute the bottleneck distance between the persistence diagram of the lazy witness filtration and that of the Vietoris-Rips filtration induced by the $\epsilon$-net landmarks for different values of $\epsilon$. For each $\epsilon$ and algorithm, we generate $10$ sets of $\epsilon$-net landmarks, compute their corresponding persistence diagrams and plot the mean and standard deviation of the bottleneck distances in Figure~\ref{fig:torus_vali}. Since the theorem is valid for $\alpha \geq 2\epsilon$, we exclude the homology classes born below $2\epsilon$ before the distance computation. The algorithms satisfy the bound as the distances are always less than the theoretical bound of $3\log{3}$. Since the plots on the other datasets support these claims, for the sake of brevity, we omit them.
\subsection{Effectiveness and Efficiency of Algorithms Constructing $\epsilon$-nets}
\begin{figure}[p]\vspace*{-1.4em}
     \centering
     \begin{subfigure}[b]{0.8\textwidth}
         \centering
         \includegraphics[width=\textwidth,height=0.3\textheight]{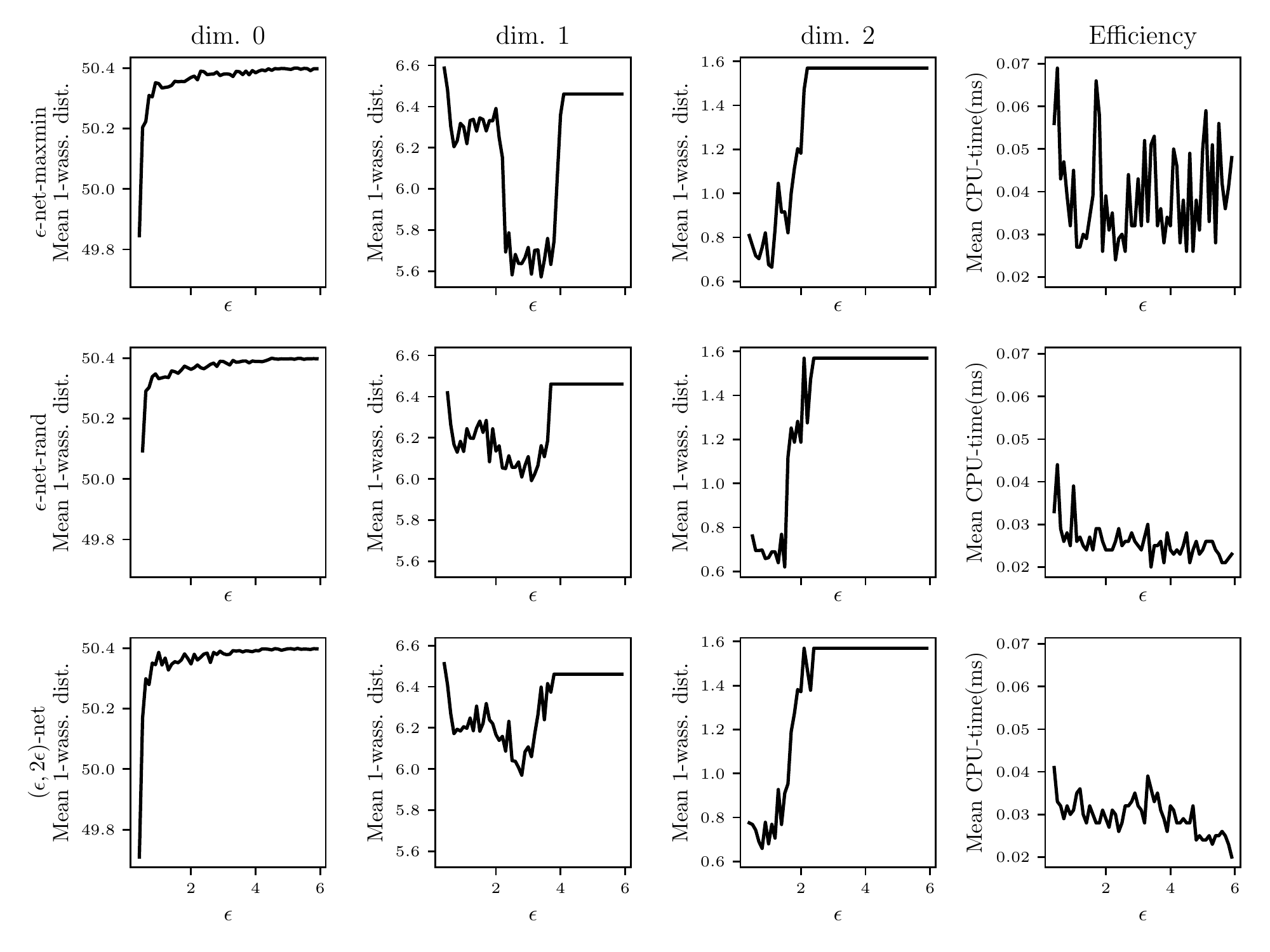}\vspace*{-1em}
         \caption{\label{sfig:torus_enet}Effectiveness and Efficiency of the $\epsilon$-net algorithms on Torus dataset.}
     \end{subfigure}
     \begin{subfigure}[b]{0.8\textwidth}
         \centering
         \includegraphics[width=\textwidth,height=0.2\textheight]{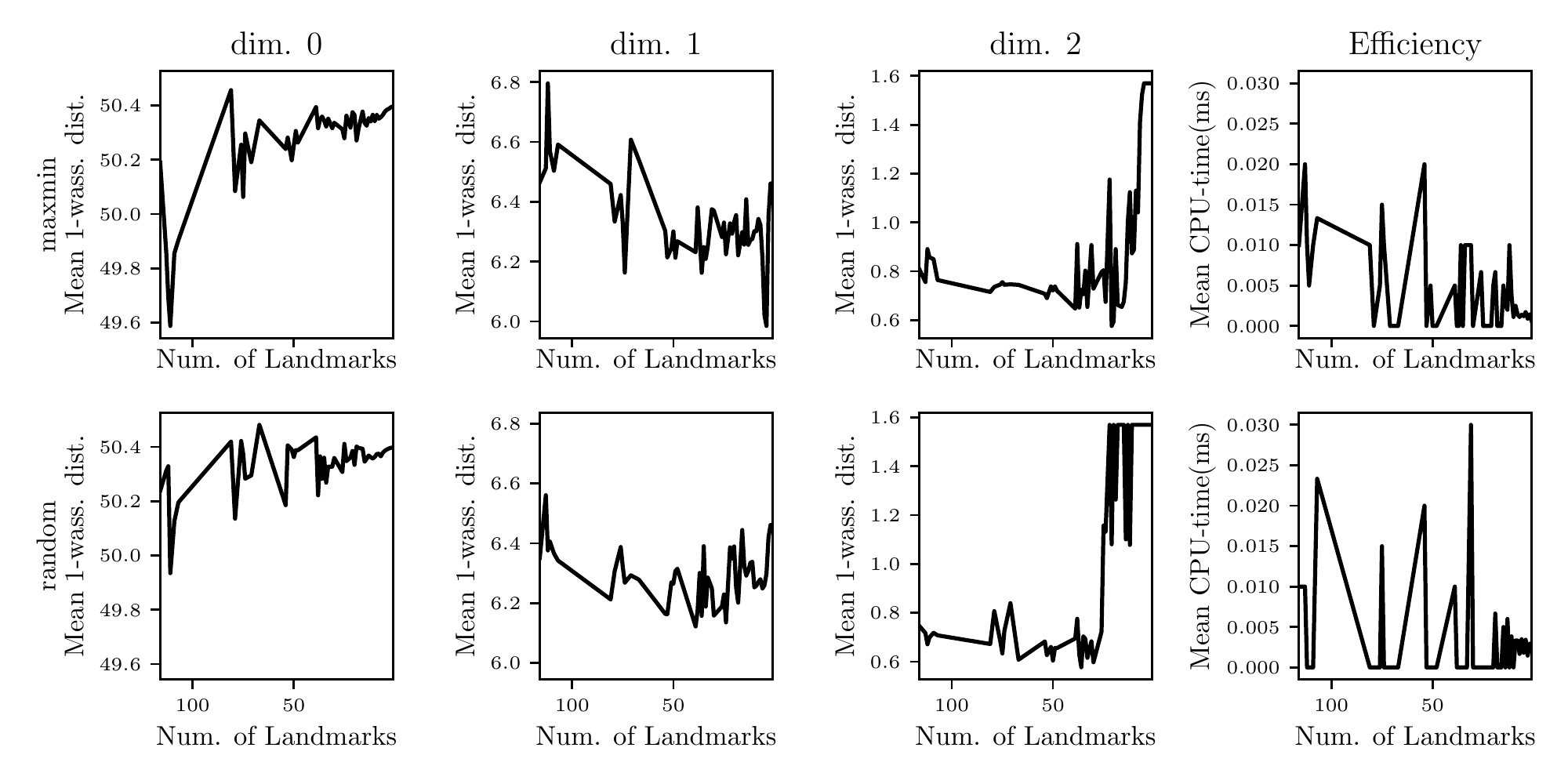}\vspace*{-1em}
         \caption{\label{sfig:torus_others}Effectiveness and Efficiency of the existing algorithms on Torus dataset.}
     \end{subfigure}
     \caption{\label{fig:tor_ef}Torus dataset.}
\begin{subfigure}[b]{0.9\textwidth}
         \centering
         \includegraphics[width=\textwidth,height=0.25\textheight]{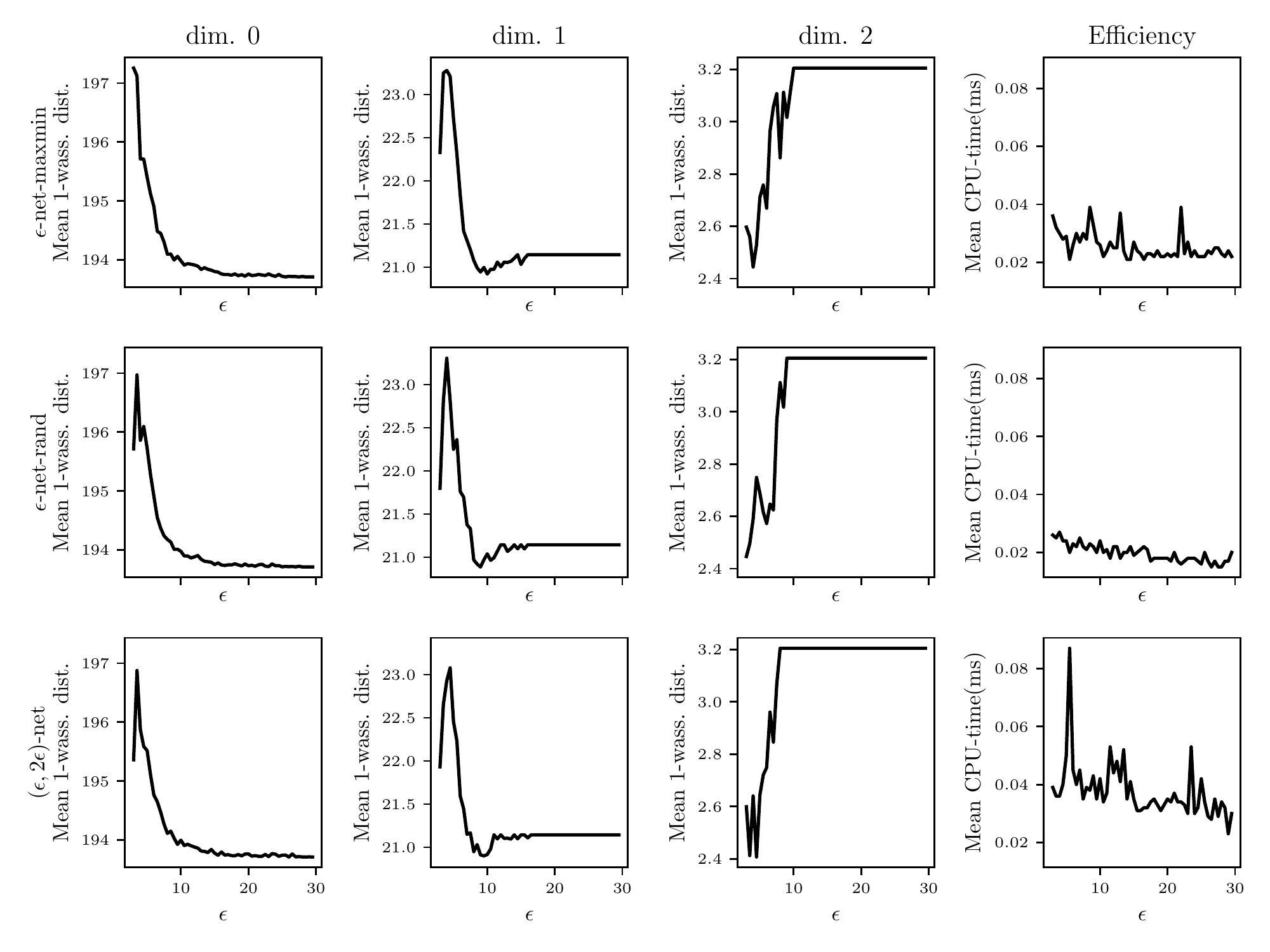}\vspace*{-1em}
         \caption{\label{sfig:grm_enet}Effectiveness and Efficiency of the $\epsilon$-net algorithms on 1grm dataset.}
     \end{subfigure}
     \begin{subfigure}[b]{0.9\textwidth}
         \centering
         \includegraphics[width=\textwidth,height=0.2\textheight]{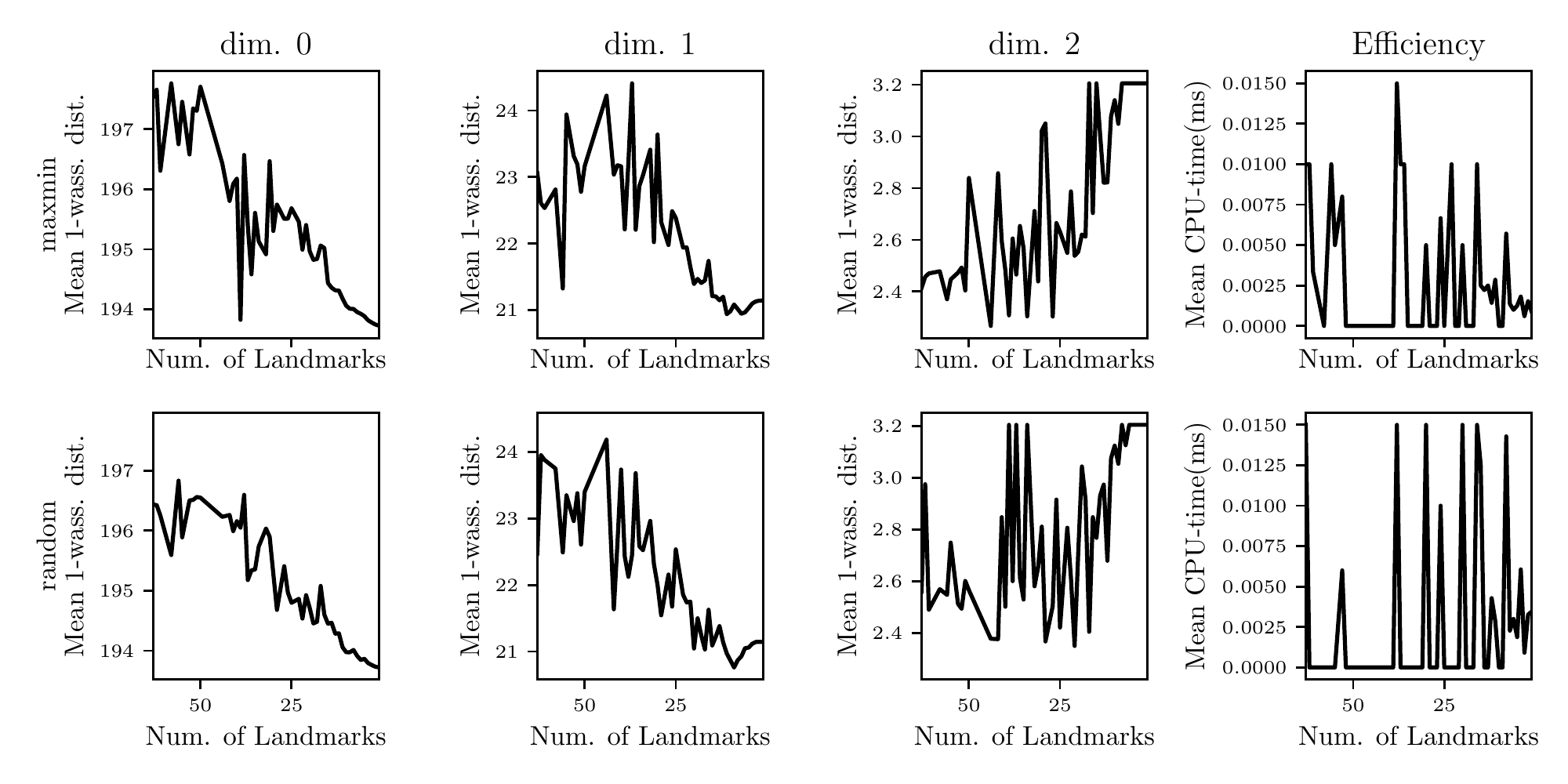}\vspace*{-1em}
         \caption{\label{sfig:grm_others}Effectiveness and Efficiency of the existing algorithms on 1grm dataset.}
     \end{subfigure}
     \caption{\label{fig:grm_ef} 1grm dataset.}
\end{figure}

For each $\epsilon$, we compute the 1-Wasserstein distance between the persistent diagrams of the lazy witness filtration induced by each $\epsilon$-net landmarks and that of the Vietoris-Rips filtration induced by the whole point cloud. We compute the mean distance and mean CPU-times across $10$ runs. Unlike $\epsilon$-net algorithms, the existing landmark selection algorithms take the number of landmarks as input. Since the average number of landmarks selected by the $\epsilon$-net algorithms does not vary much across different algorithms (Figure~\ref{fig:torus_rel}), for each $\epsilon$, we take the same number of $\epsilon$-net-maxmin landmarks as parameters to select the random and maxmin landmarks. Figure~\ref{fig:tor_ef} illustrates result on Torus dataset.

We observe that maxmin performs well in dimensions $0$ and $2$ whereas $(\epsilon,2\epsilon)$-net has competitive effectiveness.  In dimension $1$, we observe that $\epsilon$-net-maxmin achieves the lowest minimum, whereas random achieves the highest minimum. All the $\epsilon$-net algorithms has two local minima, the first of which at around $\epsilon= 0.5$ and the second in between $\epsilon=2$ to $\epsilon=4$. The first local minimum is due to the minor radius. As for the explanation of the second local minimum, it is sufficient to either cover the inner diameter of $5$ or the outer diameter of $6$ to capture the cycle. A $2.5$- to $3$-sparse sample suffices to do so. The performance of the maxmin and random landmarks is not as explainable as the $\epsilon$-net landmarks. 
In terms of efficiency, we observe that that $(\epsilon,2\epsilon)$-net algorithm has the lowest run-time among all the $\epsilon$-net algorithms. 
The $(\epsilon,2\epsilon)$-net algorithm has competitive effectiveness and better efficiency among the proposed algorithms. 
\begin{figure}[t!]
\centering
\includegraphics[height=0.5\textheight, width=\textwidth]{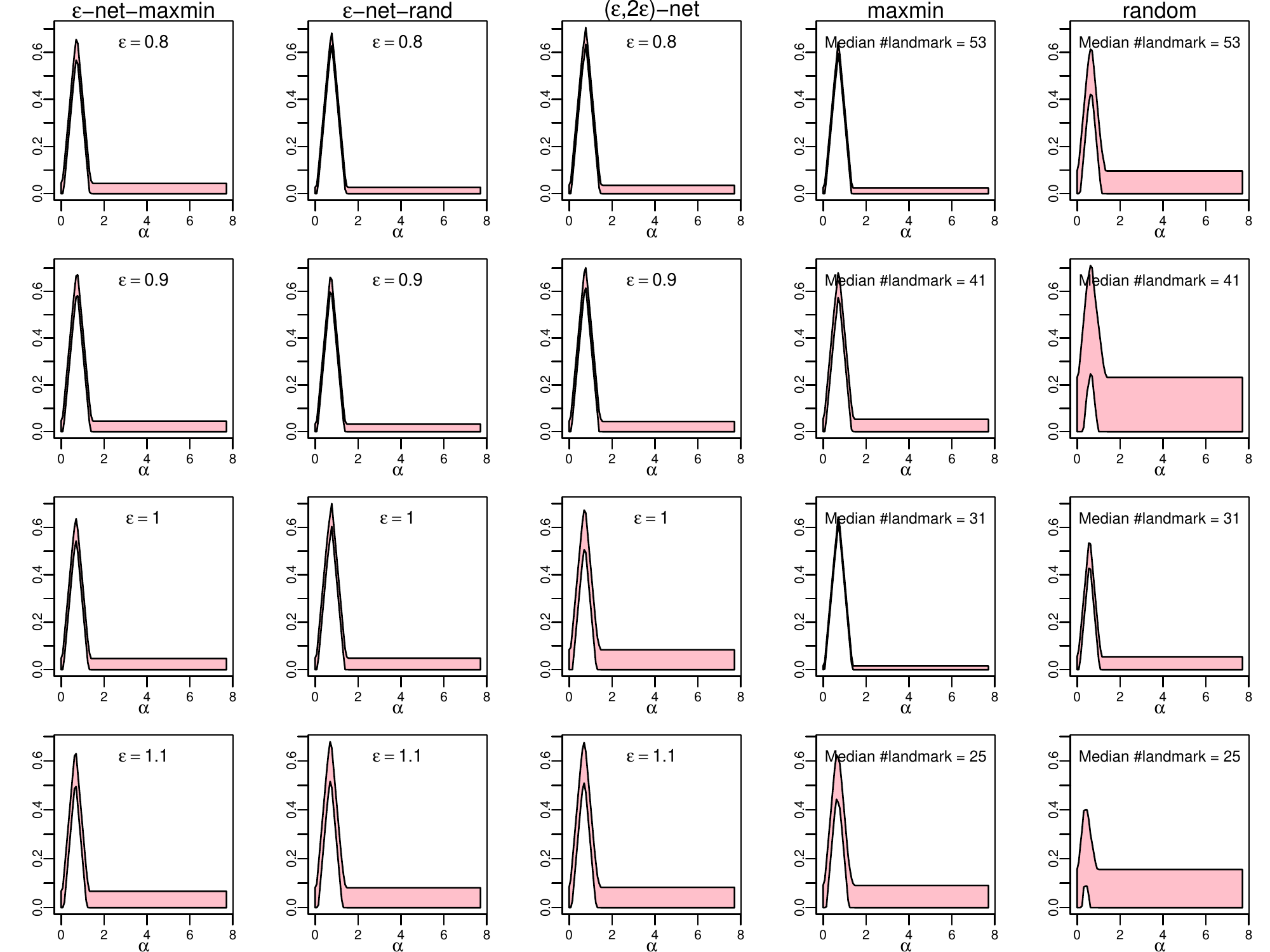}\vspace*{-.5em}
\caption{\label{fig:ttorus_stab} $95\%$ confidence band of the rank one persistence landscape at dimension 1 of the lazy witness filtration induced by the landmark selection algorithms on Tangled-torus dataset.}
\end{figure}

Figure~\ref{fig:grm_ef} illustrates the results for 1grm dataset. We observe that $(\epsilon,2\epsilon)$-net achieves the smallest loss in dimensions $0$ and $1$. In dimension $2$, maxmin achieves the smallest loss. $\epsilon$-net-rand takes the smallest CPU-time among all the $\epsilon$-net algorithms. We observe that the effectiveness of $(\epsilon,2\epsilon)$-net and efficiency of $\epsilon$-net-rand in the results on Tangled-torus dataset. We omit the plots due to space limitation. 

Despite providing better efficiency and equivalent effectiveness on the datasets under study, the performance of the  maxmin algorithm is less predictable and less explainable than the $\epsilon$-net algorithms. Among the $\epsilon$-net algorithms, $(\epsilon,2\epsilon)$-net has better effectiveness at the expense of little loss in efficiency, whereas $\epsilon$-net-rand has better efficiency than the others with effectiveness comparable to $\epsilon$-net-maxmin.

\subsection{Stability of the $\epsilon$-net Landmarks}
In Figure~\ref{fig:ttorus_stab}, we vary $\epsilon$ and plot the rank 1 persistence landscape at dimension $1$ and its $95\%$ confidence band corresponding to the lazy witness filtration induced by the different landmark selection algorithms. For maxmin and random, we take the same number of landmarks as that in the corresponding $\epsilon$-net-maxmin landmarks. The rank 1 persistent landscape is a functional representation of the most persistent homology class, which we observe from Figure~\ref{fig:ttorus_stab} in the form of a peak for all the algorithms. The $x$-axis represents the value of filtration parameter and $y$-axis represents function values.
We observe that the $\epsilon$-net-maxmin has similar confidence bands as maxmin, whereas the confidence bands of $\epsilon$-net-rand and $(\epsilon,2\epsilon)$-net are often narrower than both maxmin and random. Random has the widest confidence band among all. The confidence bands of maxmin are in-between these two extremes. This observation implies that the $\epsilon$-net algorithms are more stable than the existing algorithms. We observe similar stability results on other datasets that we omit due to space limitation. 
\section{Conclusion}
\label{sec:conc}
We use the notion of $\epsilon$-net to capture bounds on the loss of the topological features of the induced lazy witness complex. We prove that $\epsilon$-net is an $\epsilon$-approximation to the original point cloud and the lazy witness complex induced by $\epsilon$-net is a $3$-approximation to the Vietoris-Rips complex on the landmarks for values of filtration parameter beyond $2\epsilon$. Such quantification of approximation for lazy witness complex was absent in literature and is not derivable for algorithms limiting the number of landmarks.

We propose three algorithms to construct $\epsilon$-net landmarks. We show that the proposed $\epsilon$-net-rand and $\epsilon$-net-maxmin algorithms are variants of the algorithm random and maxmin respectively, which ensures the constructed landmarks to be an $\epsilon$-sample of the point cloud. 
We empirically and comparatively show that the sizes of the landmarks that our algorithms construct agree with the bound on the size of $\epsilon$-net. We empirically validate our claim on the topological approximation guarantee by showing that beyond $2\epsilon$ filtration value, the bottleneck distances are bounded by $3\log{3}$. Furthermore, we empirically and comparatively validate the effectiveness, efficiency and stability of the proposed algorithms on representative synthetic point clouds as well as a real dataset. Experiments confirm our claims by showing equivalent effectiveness of the algorithms constructing $\epsilon$-net landmarks at the cost of a little decrease in efficiency but offering better stability.\vspace*{-.1em}

\section*{Acknowledgement}This work is supported by the National University of Singapore Institute for Data Science project WATCHA: WATer CHallenges Analytics.
\bibliographystyle{splncs03}
\bibliography{main} 
\end{document}